\documentclass[%
 reprint,
 amsmath,amssymb,
prb,
]{revtex4-1}

\usepackage{graphicx}% Include figure files
\usepackage{dcolumn}% Align table columns on decimal point
\usepackage{bm}% bold math
\usepackage{xcolor}
\begin{document}

\preprint{APS/123-QED}

\title{On the performance of a photosystem II reaction centre-based photocell}% Force line breaks with \\

\author{Richard Stones}
\affiliation{%
 Department of Physics and Astronomy, University College London, Gower Street, London, WC1E 
}%
\author{Hoda Hossein-Nejad}
\affiliation{%
 Department of Physics and Astronomy, University College London, Gower Street, London, WC1E 
}%
\author{Rienk van Grondelle}
\affiliation{%
 Department of Physics and Astronomy, VU University, 1081 HV Amsterdam, The Netherlands
}%
\author{Alexandra Olaya-Castro}%
 \email{a.olaya@ucl.ac.uk}
\affiliation{%
 Department of Physics and Astronomy, University College London, Gower Street, London, WC1E 
}%
\date{\today}% It is always \today, today,

\begin{abstract}
The photosystem II reaction centre is the photosynthetic complex responsible for oxygen production on Earth. Its water splitting function is particularly favoured by the formation of a stable charge separated state via a pathway that starts at an accessory chlorophyll. Here we envision a photovoltaic device that places one of these complexes between electrodes and investigate how the mean current and its fluctuations depend on the microscopic interactions underlying charge separation in the pathway considered. Our results indicate that coupling to well resolved vibrational modes does not necessarily offer an advantage in terms of power output but can lead to photo-currents with suppressed noise levels characterizing a multi-step ordered transport process. Besides giving insight into the suitability of these complexes for molecular-scale photovoltaics, our work suggests a new possible biological function for the vibrational environment of photosynthetic reaction centres, namely, to reduce the intrinsic current noise for regulatory processes.
\end{abstract}

\maketitle

Life on Earth is fueled by photosynthesis, the process by which plants, algae and certain bacteria convert solar energy into stable chemical energy \cite{Blankenship2001}. The initial electron transfer steps during solar energy conversion by these organisms take place in photosynthetic reaction centres (PRCs), sophisticated trans-membrane supramolecular pigment-protein complexes that exhibit a dual device-like functionality. Under illumination, a PRC complex effectively operates as Nature's solar cell\cite{Blankenship2001} where electronic excitations of chromophores are transformed into stable charge-separated states, with electron donor and electron acceptor separated by a few nanometres. This picosecond charge separation process occurs with near unit quantum efficiency implying that almost every quanta of energy absorbed results in charge separated across the PRC \cite{Blankenship2011,Wientjes2013}. The same chromophore-protein structure and energetic landscape promoting this quantum yield also favours a diode-like behaviour of all PRCs such that under an appropriate applied bias, electric current flows almost entirely in one direction \cite{Kamran2015}. Their functional versatility, nanometre size, and near-unit quantum efficiency has motivated the exploration of PRCs 
as possible components of photovoltaic and photoelectrochemical cells \cite{Yehezkeli2014,Gratzal2001} as well as in biomolecular electronics\cite{Kamran2015,Reiss2007,Mikayama2006}.

A step further in this field is the recent development of single-molecule techniques that allow measurement of the photocurrent through individual PRC complexes\cite{Gerster2012}. Using cysteine group mutations, it has been possible to bind a photosystem I unit to a gold substrate and use a scanning probe gold-tip that acts as both an electrode and a localized light source to excite and measure the photocurrent of a fully functional PRC\cite{Gerster2012}. Moreover, using a tapping mode atomic force microscope, it has been possible to confirm that electrons tunneling through a bacterial PRC,  under an applied bias, follow the transfer pathway of the photochemical charge separation \cite{Kamran2015}. These experiments open up a new platform to carry out further investigations on how the microscopic mechanisms underlying the function of PRCs affect the electric current output of a single PRC, as well as to reveal further details of such microscopic mechanisms.  In particular, it is foreseeable that besides measuring the current-voltage features these techniques may allow characterization of current fluctuations and the associated counting statistics of electron transport in PRCs. In quantum transport setups\cite{Blanter2000,Kiesslich2007,Ubbelohde2012} it has been shown that such fluctuations can reveal intrinsic dynamical features of the quantum system through which electron transport occurs, including the influence of electron-phonon interactions \cite{Haupt2006} and coherence  \cite{Belzig2005}. A theoretical study along the lines of counting statistics for light-harvesting complexes has been carried out\cite{HosseinNejad2013} but so far there has been no investigation of full counting statistics of charge transport in PRCs. 

The photosystem II reaction centre (PSIIRC), present in higher plants, algae and cyanobacteria\cite{Blankenship2001}, is arguably the most important prototype to be considered as it is responsible for water splitting and production of all oxygen on Earth \cite{Rutherford2003}. Experimental evidence indicates that one of the distinguishing features of PSIIRC is the existence of at least two different charge separation pathways, one of which starts in the monomeric chlorophyll of the active D1 branch (Chl$_{D1}$) \cite{Groot2005,Holzwarth2006,Romero2010}. This transfer pathway has been argued to be a deciding factor for the functional operation of PSIIRC as a water splitting complex \cite{Renger2011} and is therefore the focus of this paper. While the detailed quantum mechanical features underpinning charge separation in PRCs are still under scrutiny \cite{Renger2011}, there is a wealth of steady-state and time-resolved spectroscopy revealing the electronic state space and spectral density of fluctuations relevant for the formation of stable charge separated states \cite{Novoderezhkin2005,Novoderezhkin2004,Romero2010,Renger2011}.
However, the implications of these microscopic mechanisms for the electric current output of a single PSIIRC are largely unknown.  In fact, it is unclear whether these natural light-to-charge converting units are well suited for anthropogenic use: would photocells integrating the microscopic mechanisms of PRCs deliver optimal power? How do these mechanisms affect the statistics of electron transport? The answers to these questions hold the potential to provide valuable insight both on the biophysics of these systems and for the development of the next generation of bio-inspired energy technologies.

In this work we address these questions by envisioning a photocell device where a single PSIIRC using the Chl$_{D1}$ pathway is placed between two electrodes and investigate how the microscopic mechanisms underlying the photochemical charge separation affect the electric current output and its fluctuations, under continuous incoherent illumination. By comparing the photocell operation under different spectral densities characterizing  the interaction between electronic and vibrational degrees of freedom, we show that selective coupling to underdamped vibrations does not necessarily offer an advantage in terms of current and power output for this photocell but they lead to output currents with suppressed noise levels as quantified by a Fano factor less than one. A structured spectral density that includes coupling to well-resolved vibrations allows the noise strength to probe the structure of the exciton manifold which transfers population to charge transfer states and leads to a sub-Poissonian statistics. This indicates that both the exciton manifold and the electron-vibration interactions in PSIIRC support a multi-step ordered electron transport process. Our work therefore puts forward a new possible functional role for the vibrational environment of PRCs, that of guaranteeing intrinsic current noise control for regulatory purposes. Our analysis also gives insight into the suitability of PSIIRCs and their microscopic principles for photovoltaic or nano-electronic applications.

\section*{Results}

\begin{figure*}[ht]
\centering
\includegraphics[scale=0.6]{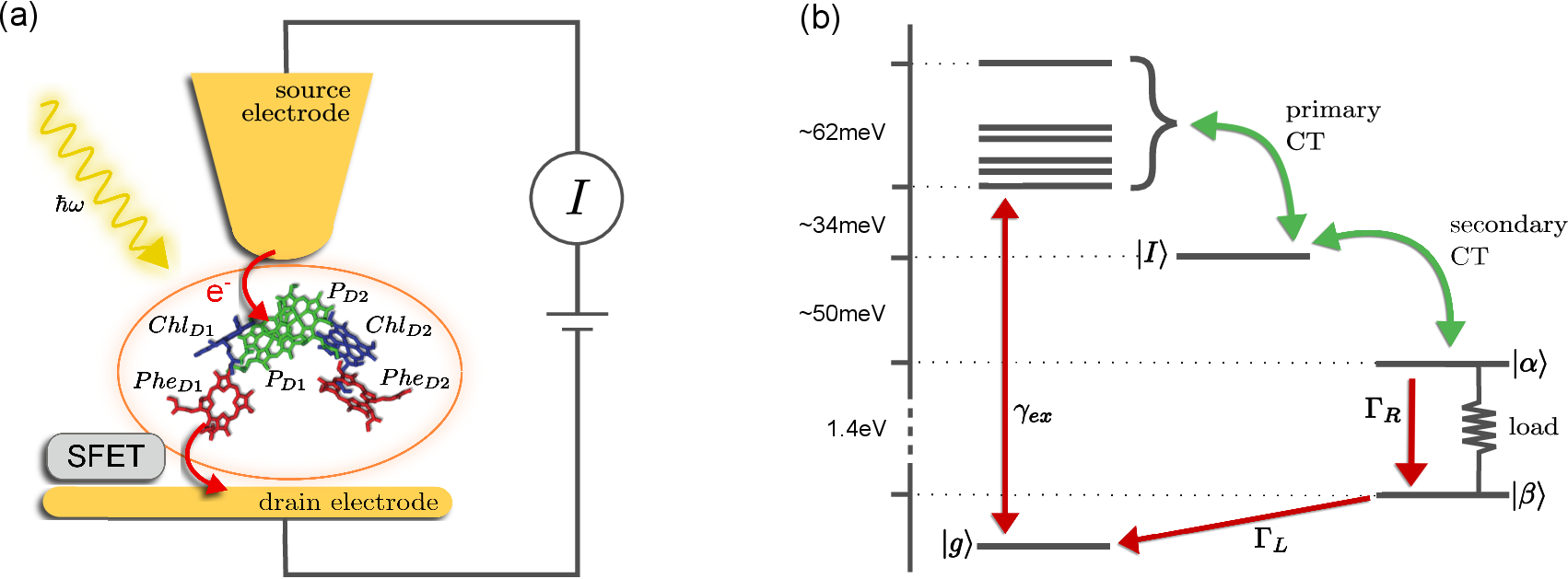}
\caption{\textbf{Photosystem II reaction centre photocell model.} (a) Schematic diagram of a proposed experimental setup for the photocell unit. The isolated core chromophores of PSIIRC are positioned between a gold substrate and a gold coated scanning probe microscope tip which act as electrodes. A silicon field-effect transistor (SFET) placed near the drain electrode could be used to measure the current statistics. (b) Energy level diagram showing the electronic state space of the model. The red arrows represent rates connecting the ground and empty state to the excited state manifold. $\Gamma_L$ and $\Gamma_R$ connect the system to the source and drain leads respectively while $\gamma_{ex}$ represents a coupling to an optical field which excites the system from the ground state to the lowest energy exciton state. Green arrows represent F\"{o}rster/Marcus rates for primary and secondary charge separation. The load between states $\alpha$ and $\beta$ indicates the transition across which we calculate the output current of the photocell and its statistics.}
\label{fig:photocell}
\end{figure*}

\begin{figure}[ht]
\centering
\includegraphics[scale=0.31]{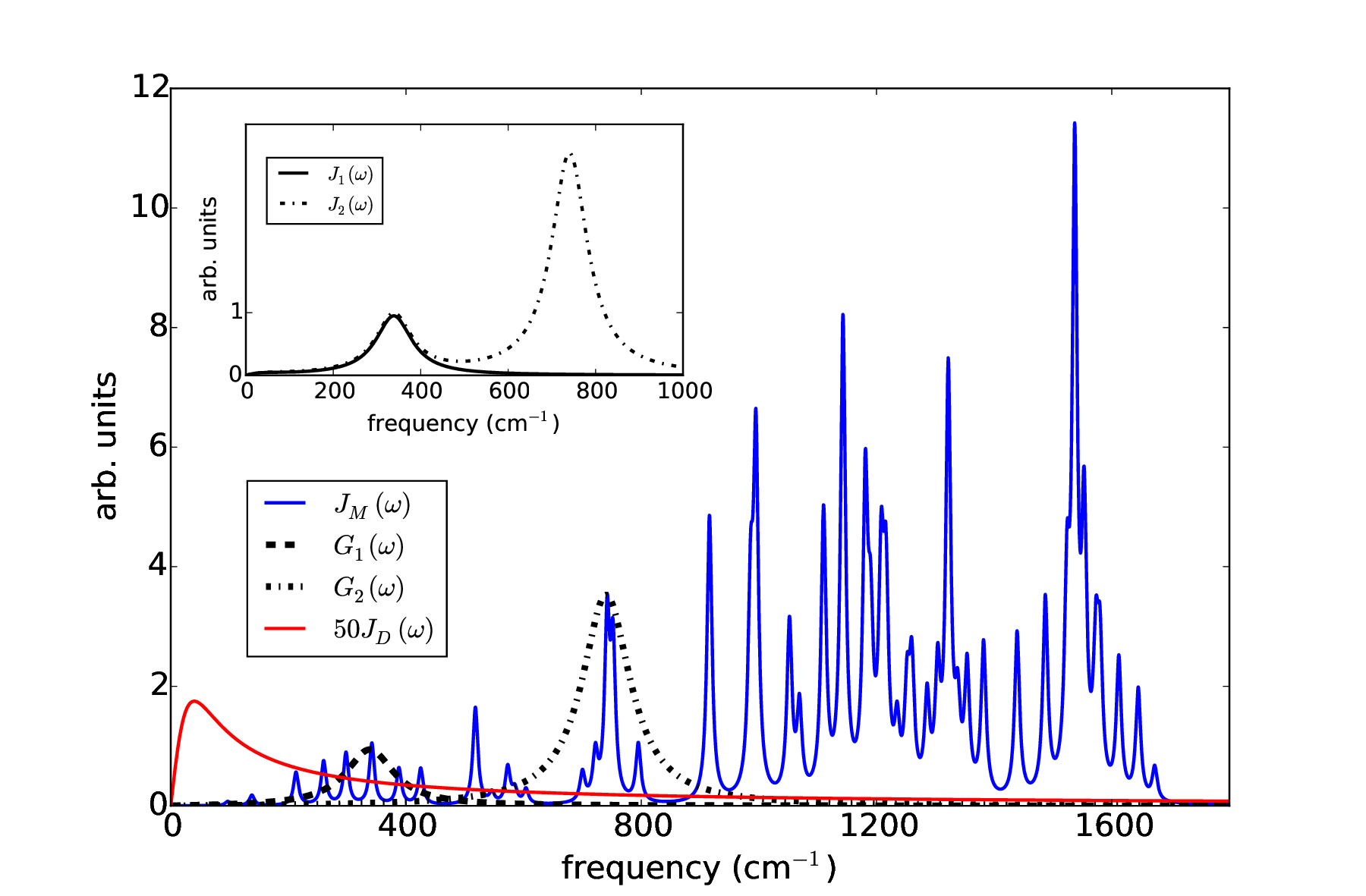}
\caption{\textbf{Photosystem II reaction centre spectral densities.} The components of the spectral densities used in the PSIIRC photocell model. Mode parameters of the structured component $J_M(\omega)$ are shown in Table S1. The Drude part $J_D(\omega)$ is scaled relative to the high energy parts for clarity.  The inset shows the spectral densities $J_1(\omega)$ and $J_2(\omega)$  which are used to approximate the full spectral density.}
\label{fig:spectral-densities}
\end{figure}

\subsection*{A photocell based on the photosystem II reaction centre}

The prototype complex we consider is the PSIIRC for which crystallography has provided the arrangement of the chromophores involved in primary charge transfer \cite{Umena2011}. As illustrated in Fig. \ref{fig:photocell} (a), four chlorophylls (Chl) and two pheophytins (Phe) are arranged in two branches (D1 and D2), where D1 and D2 label the chlorophyll binding proteins in the core of the reaction centre. The two central chlorophylls P$_{D1}$ and P$_{D2}$ (known as the special pair) are flanked by the accessory chlorophylls Chl$_{D1}$ and Chl$_{D2}$, and the pheophytins Phe$_{D1}$ and Phe$_{D2}$. Charge separation only occurs down the D1 branch \cite{Diner2002,Steffen1994}.  Nonlinear spectroscopy has revealed at least two different excited states, 
(P$_{D1}$P$_{D2}$Chl$_{D1})^{*}$ and (Chl$_{D1}$Phe$_{D1}$)$^{*}$,  that give rise to two
 different pathways (denoted the P$_{D1}$ and Chl$_{D1}$ pathways respectively) for charge separation along the D1 branch  \cite{Romero2010,Novoderezhkin2005,Novoderezhkin2011}. The likelihood of each depends on the specific protein configuration and the corresponding disorder of pigment excitation energies \cite{Romero2010}. 
 
Although the relative contribution of these pathways in ensemble measurements is not conclusively known, 
spectroscopy and its corresponding theoretical fit indicate that electron transfer is predominantly initiated from the state (Chl$_{D1}$Phe$_{D1}$)$^{*}$ \cite{Novoderezhkin2011}. This transfer pathway is considered an important building block for the functional operation of PSIIRC because by placing the Chl$_{D1}$ as the lowest energy pigment it encourages charge transfer through D1 and by lifting the energy of the P$_{D1}$P$_{D2}$ pair it provides favorable conditions for the water oxidation function \cite{Renger2011,Renger2008}. Our analysis therefore focuses on this Chl$_{D1}$ pathway in which inter-pigment electronic couplings lead to formation of delocalised exciton states upon photo-excitation as revealed by spectroscopy\cite{Novoderezhkin2007,Romero2010}. However, for this protein configuration there is no evidence of coherent coupling between excited states and charge transfer (CT) states \cite{Romero2010, Novoderezhkin2005, Novoderezhkin2011}, unlike the case of the P$_{D1}$ pathway where ultrafast non-linear spectroscopy has given evidence of coherent electron transfer\cite{Romero2010,Romero2014}. This is consistent with a very weak electronic coupling between the low energy Chl$_{D1}$ and Phe$_{D1}$ pigments and the primary  CT state (see Table S2). The Chl$_{D1}$ pathway can thus be seen as the relaxation dynamics in the energy landscape illustrated in Fig.  \ref{fig:photocell}(b). Population of an exciton state with the largest amplitude in the pair (Chl$_{D1}$Phe$_{D1}$)$^{*}$ undergoes a dynamics within the exciton manifold while it is incoherently channeled towards an initial CT state $|\textrm{Chl}_{D1}^{+}\textrm{Phe}_{D1}^{-}\rangle$ and from there to the stable secondary CT state $|\textrm{P}_{D1}^{+} \textrm{Phe}_{D1}^{-}\rangle$ in which electron and hole reside on different chromophores separated by a few nanometres along the D$1$ branch (see Fig. \ref{fig:photocell}(a)).

Fluorescence line narrowing experiments \cite{Peterman1998} have given evidence of a highly structured spectral density characterizing the interactions of an excited chromophore in PSIIRC with a wide range of vibrational motions as shown in  Fig. \ref{fig:spectral-densities}. Although some of the sharp modes in this spectral density have frequencies that match energy gaps between exciton states \cite{Romero2014}, we will show that this feature is not entirely determinant for the mean current output of a PSIIRC-based photocell operating within the Chl$_{D1}$ pathway. Together with the state-space aforementioned, this spectral density of fluctuations has provided a good fit for steady-state and transient spectroscopy of primary charge separation in PSIIRC \cite{Novoderezhkin2005,Novoderezhkin2011,Romero2014}.

We therefore consider these features of the Chl$_{D1}$ charge separation pathway to put forward two key questions: (i) \textit{do these naturally occurring electron-vibration interactions and their associated spectral density provide the best strategy for maximizing current output in a bio-inspired photocell?} and (ii) \textit{how do the exciton manifold and the vibrational environment affect the statistics of electrons flowing through this PRC?}

These questions are addressed by envisioning a theoretical photocell device in which a single PSIIRC unit operating with the Chl$_{D1}$ pathway is placed between two leads that can supply or take away electrons from the system, as illustrated in  Fig. \ref{fig:photocell} (a). The charge transfer cycle is such that electrons are pumped from the source lead at rate $\Gamma_L$ and leave the system from the drain lead at rate $\Gamma_R$ while the sample is incoherently photo-excited at rate $\gamma_{ex}$. Our model enforces the Coulomb blockade regime such that the probability of two electron occupancy of the photocell is negligible\cite{Bagrets2003}. We also consider an infinite applied  bias that assures unidirectional electron flow\cite{Marcos2010, Harbola2006}. Assuming low enough excitation rates to guarantee that only single excitation states are populated, the state space of the PSIIRC-based photocell, shown in  Fig. \ref{fig:photocell} (b), spans the following: the ground state $|g\rangle$, six exciton states $|X_1\rangle$ to $|X_6\rangle$, the initial CT state $|\mbox{Chl}^+_{D1}\mbox{Phe}^-_{D1}\rangle \equiv |I\rangle$, the secondary CT state $|\mbox{P}^+_{D1}\mbox{Phe}^-_{D1}\rangle \equiv |\alpha\rangle$ and the positively charged state $|\mbox{P}^+_{D1}\mbox{Phe}_{D1}\rangle \equiv |\beta\rangle$ which represents the `empty' state of the system for counting statistics calculations. The six exciton states arise from diagonalization of the site part of Hamiltonian $H_{el}$ (see Methods) which includes coherent electronic interactions among all the six core chromophores located in both the D1 and D2 branches of the PSIIRC. Although the D2 branch is not directly involved in charge separation, excitons localized here can act as electronic traps \cite{Novoderezhkin2011} and therefore can affect the statistics of electrons flowing through the system as we will discuss later.

To investigate the effects of the vibrational environment in the performance of the photocell, we aim to compare four cases corresponding to the four different spectral densities depicted in Fig. \ref{fig:spectral-densities}: (i) the case where the full structured spectral density $J(\omega)= J_{D}(\omega)+ J_M(\omega)$ is considered, the cases where, besides the low-energy background $J_D(\omega)$, we account for one and two well-resolved modes with spectral densities (ii) $J_{1}(\omega)=J_D(\omega)+G_1(\omega)$ and (iii) $J_{2}(\omega)=J_1(\omega) + G_2(\omega)$, and (iv) the case where only the smooth low energy component $J_{D}(\omega)$ is included. The expressions for the different components are given by\cite{Novoderezhkin2005}: 
\begin{eqnarray}
J_{D}(\omega) &=& \frac{2\lambda_D\omega\Omega_D}{\omega^2 + \Omega_D^2}, \\
G_{j}(\omega) &=& \frac{2\lambda_j\omega_j^2\gamma_j \omega}{(\omega^2 - \omega_j^2)^2 + \gamma_j^2\omega^2}. 
\end{eqnarray}
$J_{D}(\omega)$ is the Drude form of a spectral density describing an overdamped Brownian oscillator where $\lambda_D$ and $\Omega_D$ are the reorganisation energy and cut off frequency, respectively. $G_i(\omega)$ describes the spectral density of an underdamped mode coupled to an excited pigment, with $\lambda_j$, $\omega_j$ and $\gamma_j$ being the reorganisation energy, frequency and damping rate of mode $j$ respectively. $J_M(\omega)=\sum_j G_j(\omega)$ has been measured experimentally \cite{Peterman1998} and includes 48 underdamped modes. For case (ii) described by $J_1(\omega)$, we consider $\omega_1=342\textrm{cm}^{-1}$ and for case (iii) corresponding to $J_2(\omega)$, we consider $\omega_1$ as well as $\omega_2=742\textrm{cm}^{-1}$. These two modes have been argued to be important for electron transfer along the alternative P$_{D1}$ charge separation pathway\cite{Romero2014} and in our case their  frequencies span all the energy gaps of the system. Hence, $J_2(\omega)$ provides a good approximation to the full spectral density. All parameters for these spectral densities are detailed in Supplementary Note 1. By comparing these cases we attempt to address the question of how well `adapted' are these electron-vibrational interactions for photovoltaics: if it were possible to decouple these well resolved nuclear motions from the charge separation process, would the resultant photocell exhibit a better current and power output? Besides its theoretical relevance, this comparison is experimentally motivated as such decoupling may be feasible via optical cavities\cite{Coles2014,Herrera2016}.

As we will discuss in the next section, the mean current and power of our photocell is determined by the steady-state population of the secondary CT state $|\alpha\rangle$. A non-perturbative computation of the steady-steady state under the influence of the full spectral density $J(\omega)$ with its 48 sharp modes per electronic state is quite challenging and out of the scope of our computational capabilities. We can however compute non-perturbative dynamics and steady state of our photocell including coherent interactions among single-excitation states and under the influence of $J_D(\omega)$,   $J_1(\omega)$  and $J_2(\omega)$ using the hierarchical equations of motion \cite{Ishizaki2009,Shi2009,Tanimura2012} (see Methods). For comparison we also investigate the dynamics and steady state predicted by a simpler Pauli master equation for state populations, with transfer rates as described in Ref. \cite{Novoderezhkin2011} (see Methods). Theoretical justification of the validity of this approximate framework is presented in Supplementary Note 3. Figures S2 and S4 show that the population dynamics and steady-state populations predicted by the accurate framework and the Pauli master equation agree qualitative and quantitatively. Furthermore it has been shown that this approximate scheme accurately reproduces transient and steady-state spectroscopy of the PSIIRC \cite{Novoderezhkin2011}. The main difference we see is that, as expected, the accurate framework predicts some short-lived excitonic coherences (see Fig. S3) which we show to have negligible influence on the current and power delivered by our photocell.

\subsection*{Photocell current and power performance}

We fix the rate $\Gamma_L$ at which electrons are injected and set $\gamma_{ex}$ to simulate excitation by concentrated solar radiation\cite{Dorfman2013,Creatore2013a}, ensuring detailed balance as specified in Supplementary Note 2. The generated steady-state current passing between the system and drain electrode is equivalent to the current flowing across a hypothetical load connecting the final states $|\alpha\rangle$ and  $|\beta\rangle$, which have an associated energy gap $E_{\alpha\beta} = E_{\alpha} - E_{\beta}$. The voltage $V$ across such a load quantifies the extractable energy from our photocell with final energy gap $E_{\alpha\beta}$ and an expression for $V$ can be derived following standard thermodynamic considerations of photocells \cite{Shockley1961} and photochemical systems \cite{Ross1967}. Denoting the steady state populations of the secondary CT state $\rho_{\alpha\alpha}$ and the `empty' state $\rho_{\beta\beta}$, the load voltage $V$ can be expressed as:
\begin{equation}
eV = E_{\alpha\beta} + k_B T \mbox{ln}\left[\frac{\rho_{\alpha\alpha}}{\rho_{\beta\beta}}\right],
\end{equation}
where $k_B$ is the Boltzmann constant, $T$ the temperature of the photocell and $e$ the electric charge. The average current $\langle I \rangle$ and the power output $P$ delivered by the photocell are given, respectively,  by $\langle I \rangle = e\Gamma_R\rho_{\alpha\alpha}$ and $P = \langle I \rangle V$. By fixing all parameters except the rate $\Gamma_R$ at which electrons leave the system, we can then investigate the characteristic $\langle I \rangle-V$ and $P-V$ curves which define the photovoltaic performance of a photocell.

\begin{figure}[!t]
\centering
\includegraphics[scale=0.3,trim={1cm 0 0 0},clip]{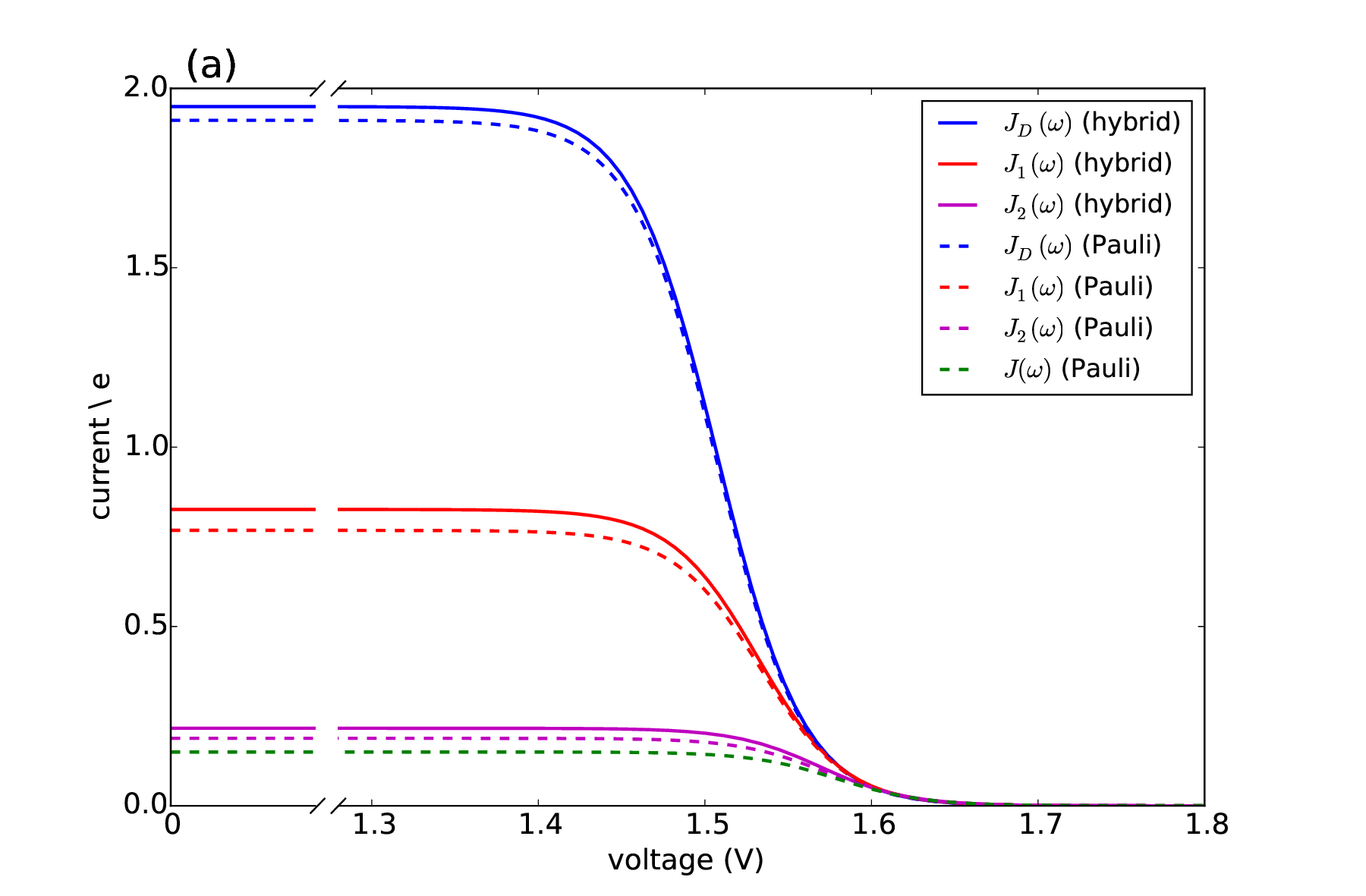}
\includegraphics[scale=0.3,trim={1cm 0 0 0},clip]{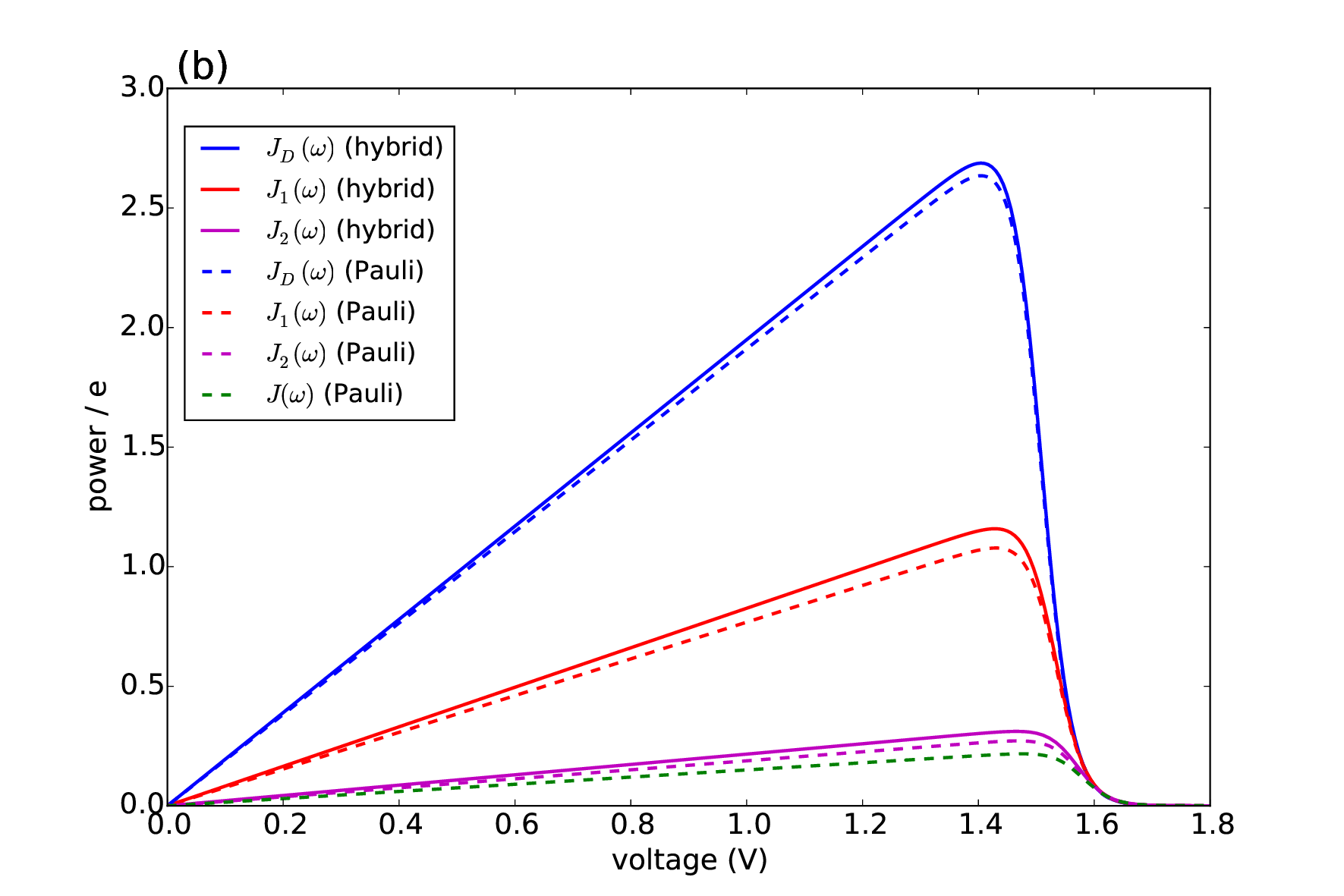}
\caption{\textbf{Photocell mean current and power versus voltage for different spectral densities.} (a) Current. (b) Power. Solid lines are calculated using HEOM hybrid model while dotted lines are calculated using the Pauli master equation. Calculations carried out at 300K with excitation rate $\gamma_{ex} = 75\mbox{cm}^{-1}$. See Supplementary Note 1 for all other parameters.}
\label{fig:current-power-voltage}
\end{figure}

Figure \ref{fig:current-power-voltage} presents the characteristic curves for the four spectral densities depicted in Fig. \ref{fig:spectral-densities}. For $J_D(\omega)$, $J_1(\omega)$ and  $J_2(\omega)$, we present the results obtained both by the hybrid framework and by the approximate Pauli master equation, showing their remarkable agreement for the current and power predictions (consistent with the dynamics shown in Figs. S2 and S4).  In all the cases, the limit of $\Gamma_R \rightarrow 0$ leads to $\langle I \rangle \rightarrow 0$ defining the maximum available voltage or open circuit regime $eV_{oc}$ which is proportional to the energy gap between the ground state and the state that is directly photo-excited \cite{Creatore2013a} i.e. $eV_{oc}\approx E_1-E_g\approx 1.8\,\textrm{eV}$ . In the opposite limit, when  $\Gamma_R \rightarrow \infty$, $V \rightarrow 0$. In these two extremes the photocell delivers no power.  For all spectral densities we observe that the current is constant at low voltages and drops off at a characteristic voltage $V$ comparable to $E_{\alpha\beta}$ when the spectral density is $J_D(\omega)$. This characteristic voltage increases slightly as the spectral density includes more well-resolved modes.  
%for $J_1(\omega)$ and $J(\omega)$.

There are two remarkable features to highlight in Fig. \ref{fig:current-power-voltage}. First, the constant current observed for small voltages and the maximum power are significantly lower for a photocell with a structured environment. We observe such a reduction even in the case of $J_1(\omega)$ whose central frequency is quasi-resonant with several exciton energy gaps. This contrasts with the power enhancement predicted for a simple light-harvesting unit operating under coherent interactions between all states \cite{Killoran2015}. Second, the current delivered by a photocell with $J_2(\omega)$ is already quite close to that of $J(\omega)$, confirming that for the performance of the photocell, $J_2(\omega)$ is a good approximation to the full spectral density.

The behaviour predicted for  the current and power delivered by the PSIIRC photocell relies on two main features characterizing the Chl$_{D1}$ charge separation pathway: the large reorganisation energies of the CT states and the concomitant weak coupling between the primary CT state and single-excitation states. These two facts lead to an incoherent population dynamics that is well described by second order perturbation theory in the electronic coupling and such that the rates of transfer between an exciton $|X\rangle$ and the primary CT state $|I\rangle$ are dominated by the convolution of their line shape functions\cite{Forster1965,Yang2002}. Specifically, the rate of transfer between these states is given by $k_{X,I} = |V_{X,I}|^2 S_{X,I}$ where $V_{XI}$ is the effective electronic coupling between the states and $S_{XI}$ quantifies the spectral overlap:
\begin{equation}
S_{XI} = 2\mathcal{R}e \int^{\infty}_0 dt e^{i\omega_{XI}t} e^{-i(\lambda_X + \lambda_I)t} e^{-(g_X(t) + g_I(t))}, 
\end{equation}
with $\omega_{XI}$ the energy gap between the states, $\lambda_{X(I)}$ the corresponding reorganisation energies for each state and $g_{X(I)}(t)$ the associated line broadening functions. Full expressions for these functions can be found in Supplementary Note 2. Figure S5 shows the overlap between the low-lying exciton $|X_1\rangle$ and $|I\rangle$ for the three spectral densities. As the spectral density contains more peaks, the reorganization energy of both states increases. However, for all the cases the values for $\lambda_{I}$ are about one order of magnitude larger than $\lambda_{X}$ (see Table S5). The larger $\lambda_{I}$, the wider is the shift between donor and acceptor states and as such the overlap $S_{XI}$ accounting for spectral resonances is reduced to yield a lower rate of transfer $k_{X,I}$. Similar considerations apply for the transfer rate between the CT states $|I\rangle$ and $|\alpha\rangle$ whose reorganisation energies satisfy  $\lambda_{I}<\lambda_{\alpha}$  as discussed in Supplementary Note 4. In this scenario the condition of quasi-resonance between electronic gaps and well-resolved vibrations becomes irrelevant for population transfer. The reduced transfer rates from excitons to the intermediate CT state  $|I\rangle$ can then be interpreted as a Zeno-like effect whereby the strongly coupled environment ``measures" the population of the CT state at a very high rate thereby slowing transfer. The disparity between reorganisation energies for CT states and the donor exciton states ensures a downhill relaxation that in the biological context counteracts charge recombination.

\subsection*{Zero-frequency noise}

\label{sec:noise}
Full performance characterization of a photocell cannot be limited to the steady-state current and power output. Steady-state current fluctuations quantified by second and  higher order cumulants ($\langle I^n\rangle_c$ with $n>1$) can give information on the microscopic mechanisms underlying correlations between elementary charge transfer events\cite{Blanter2000}. While the theory of full counting statistics is well established for Markovian systems\cite{Bagrets2003,Marcos2010}, for non-Markovian dynamics only a few comprehensive frameworks based on perturbative approaches have been put forward\cite{Flindt2010}. To address this shortcoming, we have developed a non-perturbative formalism to compute the full counting statistics \cite{Stones2017}, which integrates the exact system  dynamics provided by the hierarchical equations of motions with a recursive scheme that allows accurate computation of the current cumulants \cite{Flindt2010} (see Methods).

We focus on the long-time limit or zero-frequency regime of the relative noise strength which is quantified by the second order Fano factor \cite{Fano1947}: 
\begin{equation}
F^{(2)} = \frac{\langle I^2\rangle_c}{\langle I\rangle}.
\end{equation}
This ratio between the zero-frequency second order current cumulant and the mean current quantifies the deviation of the underlying statistical process from a Poissonian distribution. A Fano factor of 1 indicates a Poissonian process without correlations among charge transfer events. Deviations from 1 are interpreted as either super-Poissonian ($F^{(2)} > 1$) or sub-Poissonian ($F^{(2)} < 1$), regimes that can be associated with highly fluctuating or more stable currents, respectively.

\begin{figure}[!t]
\centering
\includegraphics[scale=0.4,trim={0.8cm 0 0 3.5cm},clip]{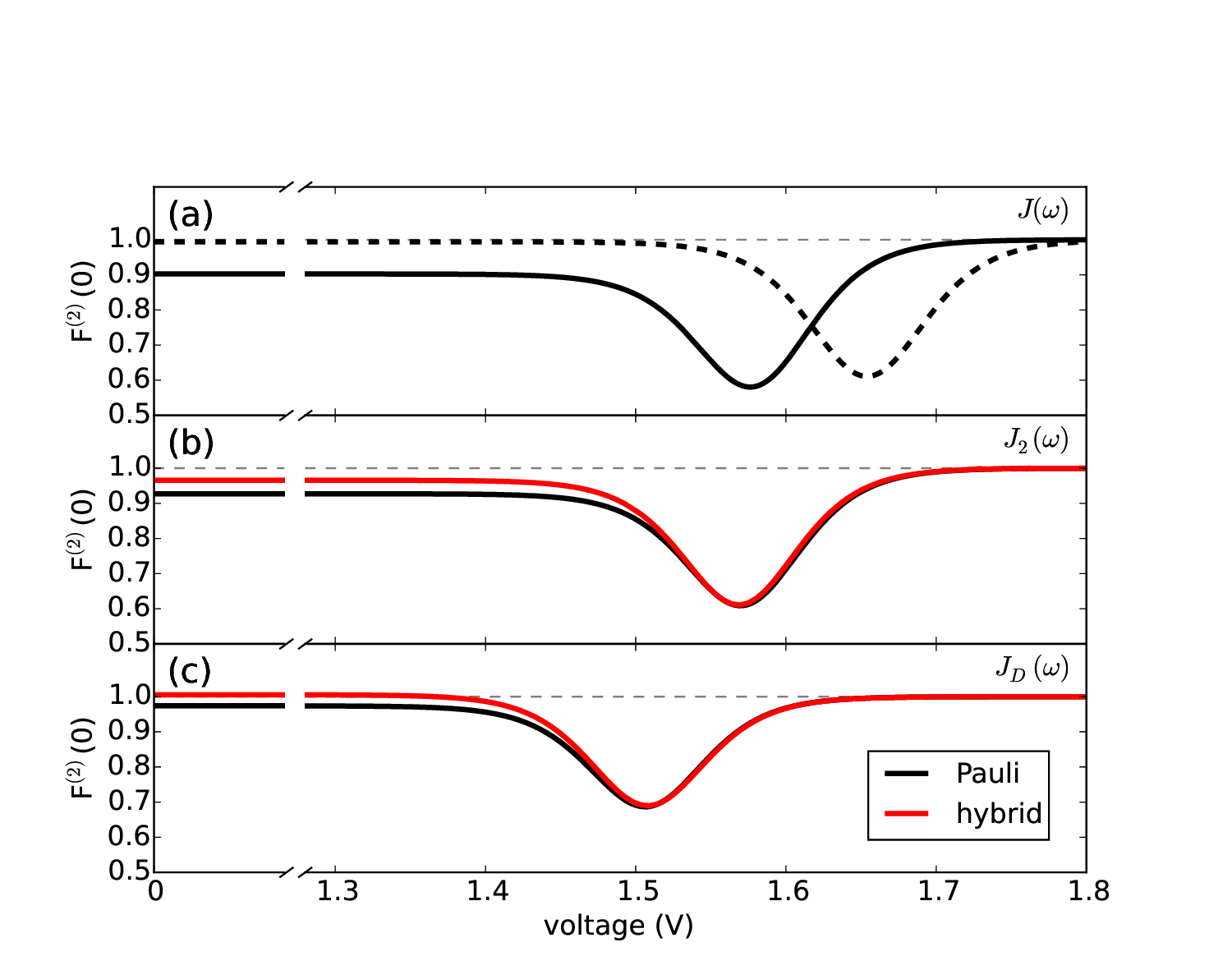}
\caption{\textbf{Fano factor versus voltage.} (a) Fano factor for the PSIIRC photocell with the structured spectral density $J(\omega)$ for a modified (slow) (dotted line) and the measured (solid line) secondary charge transfer rate. (b) Fano factor for the PSIIRC photocell with spectral density $J_2(\omega)$ containing two underdamped vibrations. (c) Fano factor for the PSIIRC photocell with a smooth low-energy vibrational environment $J_D(\omega)$. Calculations carried out at 300K with excitation rate $\gamma_{ex} = 75\mbox{cm}^{-1}$. See Supplementary Note 1 for all other parameters.}
\label{fig:noise-voltage}
\end{figure}

Figure \ref{fig:noise-voltage} reports both non-perturbative and approximate results for $F^{(2)}$ versus $V$ at room temperature for $J_D(\omega)$ (Fig. \ref{fig:noise-voltage}(c)) and $J_2(\omega)$ (Fig. \ref{fig:noise-voltage}(b)). For simplicity, we have omitted the results for $J_1(\omega)$ as they simply follow the same trend. The qualitative and quantitative agreement between these curves indicate that zero-frequency noise properties of the photocell for these spectral densities is dominated by a population dynamics that is well-captured by the approximate Pauli framework. This is also consistent with the fact that exciton coherences arising from the interaction with a slowly relaxing bath or well-resolved vibrational motions decay much faster than the time scale on which the steady state is reached as discussed in the supplementary information. These arguments then justify the use of the approximate scheme to obtain insights into the behaviour of $F^{(2)}$ for the full spectral density $J(\omega)$ as shown in Fig. \ref{fig:noise-voltage}(b).

In mesoscopic  and quantum systems the zero-frequency Fano factor has proven to be very sensitive to the structure of the state space \cite{Egues1994} as well as to system-environment interactions \cite{Haupt2006}. This is precisely what is indicated by the results for the  Fano factor of the current through our photocell device shown in Figs. \ref{fig:noise-voltage} (a) and (b), which show that the noise profiles for $J_2(\omega)$ and $J(\omega)$ each have a single minimum but are not symmetric with respect to the voltage at which this minimum occurs. In both cases we see that for $V\rightarrow 0$ we have $F^{(2)} < 1$ while in the opposite limit of $V\gg E_{\alpha,\beta}$ we obtain $F^{(2)}\rightarrow 1$. This indicates that at small voltages the electron transport is slightly correlated in all cases. As the spectral density exhibits more structure, the noise levels are overall lower. For instance, for the full spectral density we have $F^{(2)}(V=0)=0.90$ while for $J_2(\omega)$, we have  $F^{(2)}(V=0)=0.95$. Similarly, the minimum of $F^{(2)}$ reaches lower values as the spectral density acquires more structure i.e. $F^{(2)}=0.55$ for $J(\omega)$ which is less than the values obtained for $J_2(\omega)$ i.e. $F^{(2)}=0.6$ and for $J_D(\omega)$ i.e. $F^{(2)}=0.7$.  In all the cases the sub-Poissonian behaviour is a manifestation of the Coulomb blockade regime where the presence of an electron in the system prevents another one entering until the system is empty. However, the more ``ordered" transport observed for spectral densities with more well-resolved spectral features relies on the rapid population transport among excitons induced by such structured vibrational environments. The rates of transport to CT states, while reduced, are still comparable to the transfer among excitons such that the statistics of transitions from states $|\alpha\rangle$ to $|\beta\rangle$ samples the manifold of exciton states donating population to the primary CT state. This hypothesis is confirmed by our analysis of the energy scale determining the voltage at which $F^{(2)}$ is minimum in each case of Fig. \ref{fig:noise-voltage}.

The analytic form of the Fano factor for our photocell model is too cumbersome to give any insight into the conditions determining the minima in Fig. \ref{fig:noise-voltage}. In order to rationalise  the  minimum in each curve it is useful to consider the case of a single resonant level (SRL) in the infinite bias limit \cite{Marcos2010}. The dynamics of this system in the basis $\{ |\mbox{occupied}\rangle, |\mbox{empty} \rangle \}$ is governed by a Liouvillian with matrix elements $\mathcal{L}_{11} = -\mathcal{L}_{21} = \Gamma_R$ and $\mathcal{L}_{22} = -\mathcal{L}_{12} = \Gamma_L$ such that the Fano factor as a function of the voltage of a load across the occupied and empty states (Eq. (3)) takes the form
\begin{equation}
F^{(2)}(V)= \frac{1 + \exp[-2(eV - E_0)/k_B T]}{(1 + \exp[-(eV - E_0)/k_B T])^2},
\end{equation}
where $E_0$ is the energy gap between the occupied and empty states. It is simple to show that  this expression is equivalent to writing the Fano factor in terms of $\Gamma_L$ and $\Gamma_R$ as $F^{(2)}=\frac{\Gamma_L^2+\Gamma_R^2}{(\Gamma_L+\Gamma_R)^2}$ (cf. Eq. (45) in Ref.\cite{Marcos2010} and see Supplementary Note 6). Figure S2 shows that $F^{(2)}(V)$ for the SRL exhibits a single minimum, just as in our PSIIRC photocell. The minimum occurs when $V_{min}=E_0$ which is equivalent to the condition of $\Gamma_L / \Gamma_R = \rho_{\mbox{occupied}} / \rho_{\mbox{empty}} = 1$ as shown in Supplementary Note 6. In this case, however, the function is symmetric about $V_{min}$ approaching 1 at both large and small voltages and indicating that electron transfer events in these extremes are uncorrelated. Based on this, we can say that near the $V_{min}$ the noise in our PSIIRC  is approximately equal to that of an effective SRL with occupied level $|\alpha^*\rangle$, empty level $|\beta\rangle$ and renormalized energy gap $E_{\alpha \beta}^*$ that determines $V_{min}$. Denoting as $E_{jk}$ the energy gap between states $|j\rangle$ and $|k\rangle$ of our photocell, we notice that for the case of  $J_D(\omega)$ the Fano factor has a minimum for $V_{min} \approx E_{I \beta} \approx 1.50 \textrm{eV}$ while for the full spectral density $V_{min} \approx E_{X_6\beta}~\approx 1.56 \textrm{eV}$ and for the two mode spectral density $J_2(\omega)$ we have, as expected, a value in between. This indicates that in the  photocell with the full spectral density, the minimum noise samples the largest energy gap between $|\beta\rangle$ and the exciton manifold  while for $J_D(\omega)$ the noise only witnesses the energy gap up to the intermediate CT state. This is consistent with the fact that the rate of transfer among excitons in the $D1$ branch are larger for $J(\omega)$ than for the other two spectral densities.

As mentioned above, the most important feature of Figs. \ref{fig:noise-voltage} (a) and (b) is the non-symmetric profile $F^{(2)}$ with respect to $V_{min}$. We now show that the  rate limiting this asymmetric behaviour is the secondary charge transfer rate. To do this we consider the situation where the rate of secondary CT transfer $k_{I,\alpha}$ is set by hand to a very low value compared to relaxation rates within the exciton manifold and the rates between the excitons and primary CT state for the full spectral density. In this case, the population of $|\alpha\rangle$ is so slow that all the internal transfers from the exciton manifold to the primary CT can be described as a single step process i.e. there is no sampling of the exciton manifold and the Fano factor tends to 1 for small and larger voltages as shown by the dotted line in Fig. \ref{fig:noise-voltage} (a). This means that the system behaves as a SRL for all values of $V$ with renormalized $\Gamma_L^*$ (cf. Fig. \ref{fig:noise-voltage} (a) with Fig. S6). The time scale of secondary charge transfer is therefore a limiting time-scale  which can lead to a variety of phenomena as will be further explored in the next section.

To conclude, our results at room temperature indicate that the PSIIRC-based photocell with the structured vibrational environment delivers less power than a photocell with an unstructured environment, yet this is accompanied by a suppression of current fluctuations. %as indicate by a noise strength close to a minimum of 0.5 
This noise reduction is a consequence of the different features of vibration-assisted transport among excitons and transport between excitons and charge transfer states, which underlies the function of our model PSIIRC.

\section*{Discussion}

Present single-molecule technologies demonstrating the ability to manipulate and measure the photocurrent of single PRCs have motivated us to investigate how the microscopic physical mechanisms underlying the function of these complexes may affect their performance as components in a photovoltaic cell. To approach this question we have brought together biological and physical perspectives by considering a PSIIRC-based photocell in a protein configuration that is argued to favour water-splitting \cite{Renger2011} and in which charge separation is initiated at the accessory Chl$_{D1}$ \cite{Groot2005,Holzwarth2006,Romero2010}. This protein configuration is characterized by the lack of coherent delocalization between excitons and CT states \cite{Romero2010}.

Our results show that a structured environment assisting electron transfer in our model PSIIRC-based photocell device acts to reduce the current and power output in comparison to a situation where electron-vibration interactions are described by a simple smooth low-energy background function. This reduction is a manifestation of a Zeno-like effect whereby the weak donor-acceptor electronic coupling concomitant with the stronger coupling of CT states to well-resolved high energy modes lead to slower transfer rates to CT states.  These observations suggest that while PSIIRC complexes operating in the Chl$_{D1}$ pathway may favour water oxidation under \textit{in vivo} conditions, they may not necessarily be well suited for maximizing current output in single-molecule photovoltaics. Notwithstanding, the predicted reduction in the average photocurrent upon inclusion of coupling to well-resolved high-energy modes is not detrimental for the biological operation of PRCs. In the biological scenario it is more important to inhibit charge recombination and to ensure the captured energy is not wasted. In the Chl$_{D1}$ transfer pathway, stronger coupling of CT states to these well-resolved vibrational motions ensures downhill relaxation thereby helping to prevent charge recombination.

For anthropogenic purposes of obtaining the largest current out of these units regardless of its fluctuations, the best strategy then may be to decouple specific vibrational motions from electronic states. Indeed, modifications of the electron-nuclei interactions can be achieved by strong coupling of pigment-protein complex to a confined optical cavity mode \cite{Coles2014,Herrera2016}. In particular, Ref.\cite{Herrera2016} has shown that the energy exchange of electronic transitions with a strongly coupled optical mode could help suppressing reorganisation energy of the nuclei thereby increasing the rate of electron transfer reactions. Alternatively, one can select PSIIRCs operating in the P$_{D1}$P$_{D2}$ pathway where coherent delocalization across exciton and CT states has been probed \cite{Romero2010,Romero2014,Fuller2014,Romero2017} and which will lead to an enhancement as predicted in Ref. \cite{Killoran2015}.

While no advantage is obtained in terms of mean current and power output, strong coupling to well resolved vibrational modes results in a reduction of current fluctuations of our PSIIRC-based photocell. This lower noise strength obeying a sub-Poissonian statistics and the associated ordered electron transport is promoted by the exciton manifold and signals out the multi-step nature of the transport process. Preliminary calculations (not shown) for a photocell with delocalized states across excitons and CT states  indicate that such a noise reduction maybe a general feature.

From the electronic-device viewpoint, reducing any kind of noise is always a desirable feature to guarantee device resolution; this includes intrinsic noise due to the inherently probabilistic nature of the process. Hence the device functionality of this noise reduction appears to be straightforward: to improve precision in the current delivered.  More interesting is to discuss the possible advantages of such noise reduction in the biological context. It is well known that noise and its control is crucial across all scales in biology \cite{Tsimring2014,Raser2005}. For instance, it has been discussed that biochemical processes that are inherently stochastic include mechanisms to control intrinsic noise and, in particular, to reduce it for regulatory processes \cite{Raser2005}. Indeed, electron transfer events in photosynthetic reaction centres belong to a larger family of stochastic transport processes in biology, some of which have already been predicted to exhibit mechanisms suppressing fluctuations below the Poisson level \cite{Brunetti2007}. Moreover, increased complexity in biological networks has been linked to intrinsic noise reduction \cite{Cardelli2016}.  We therefore argue that, for biological function of PRCs, the coupling to well-resolved vibrations and the predicted noise reduction could indeed have a regulatory function. In these systems, the final stable CT state $|\alpha\rangle$ donates an electron to quinone B and once this reduction happens, the PRC is unable to handle an excitation during a finite time. Having single electrons delivered at regular (ordered) time intervals (with narrower fluctuations of waiting times) as opposed to randomly (Poisson-like process) could avoid wasting excitations during such overly long blocking periods.

The experimental implementation of our proposal assumes PSIIRC units that have been modified to have no quinones as has been done for the protein samples used in Ref. \cite{Romero2014}. This will ensure that attachments of electrodes to individual PSIIRC units are at the level of the electron donor and electron acceptor pigments. We envision metal-protein junctions and a scanning tip microscopy setup as those that have been realised for photosystem I (PSI) units \cite{Gerster2012} and for reaction centre-enriched purple bacterial membranes \cite{Kamran2015}. Genetic manipulation of both the oxidizing and reducing sites could allow covalent attachment of the protein to the electrodes across which the photo-current could be measured \cite{Gerster2012}. To measure the elementary transfer events from the electron acceptor site to the drain electrode we envision a device, such as a silicon field-effect transistor \cite{Nishiguchi2009,Nishiguchi2011}, capable of detecting single charges and feasible to be integrated in the scanning tip setup at room temperature as shown in Fig. 1(a).

The main limitation of isolated natural photosynthetic proteins for realistic, long-lived photovoltaic applications lies in the photodamage they experience. In PSIIRC this occurs in the D1 protein resulting in a lifetime as short as tens of minutes \cite{Brinkert2016,Edelman2008}. Emerging organic alternatives, such as the synthesis of man-made protein maquettes \cite{Lichtenstein2015} has opened the possibility of building nanometric units accurately mimicking the structure of photosynthetic systems yet displaying enhanced photostability. Merging this area with photovoltaics may unleash an unforeseen remarkable development.

From the theoretical view point a few remarks must be made. As specified in the Methods section, we assume a simplified description of the coupling between the electronic system and the leads. This neglects the possibility of the leads coupling to  localized vibronic states rather than to bare electronic degrees of freedom. The same approximation has been used in other similar systems with arbitrary electron-phonon couplings \cite{Braggio2009,Santamore2013} and has been shown to give relevant  physical insight. The full extent of this effect in our system remains to be investigated. There are also additional questions about the thermodynamic consistency of calculating the power output across with the phenomenologically modeled load as raised in Ref. \cite{Gelbwaser-Klimovsky2017}. It will therefore be important to investigate entropy production \cite{Esposito2010} for our model photocell to assess its consistency with the second law.

Finally, our work has focused on the zero-frequency noise showing that, in this case, it is dominated by the population dynamics as confirmed by our comparison between the hybrid and the approximate frameworks. An extension of our study to investigate finite-frequency noise \cite{Emary2007} could therefore be a suitable alternative to obtain signatures of quantum coherence.  More generally, current statistics measurements also potentially offer a non-invasive, single system level probe of charge transfer phenomena in a wide range of biological \cite{Xiang2015,Franco2011} and chemical systems \cite{Delor2015,Bakulin2015} ranging from charge transfer along molecular wires made from DNA strands \cite{Xiang2015} to general donor-bridge-acceptor systems \cite{Delor2015,Bakulin2015} or to unveil vibrational mechanisms for odour receptors \cite{Franco2011}.

\section*{Methods}
\subsection*{Dynamical evolution of electronic excitations}
We consider an exciton dynamics described by $H_{el}=\sum_{i} e_i |i\rangle\langle i| + \sum_{ij} T_{ij}(|i\rangle\langle j| + |j\rangle\langle i|)$ where $i$ corresponds to the basis of single-excitation states of the six core chromophores i.e. $\{ |\mbox{P}_{D1}\rangle,\, |\mbox{P}_{D2}\rangle,\, |\mbox{Chl}_{D1}\rangle,\,|\mbox{Chl}_{D2}\rangle,\, |\mbox{Phe}_{D1}\rangle,\, |\mbox{Phe}_{D2}\rangle \}$ and $e_i$ are onsite energies given in Table S1. The six eigenstates of $H_{\textrm{el}}$ are denoted as $|X_1\rangle$ to $|X_6\rangle$ with corresponding eigenenergies $E_{X_1}$ to $E_{X_6}$ in ascending order.   The electronic operators $|i\rangle\langle i|$ couple linearly with coupling $g_i$ to identical  baths of harmonic oscillators $H_I=\sum_{i,\textbf{k}} g_i |i\rangle\langle i|(b_\textbf{k}^\dag+b_\textbf{k})$. The strength of the system-bath interaction is quantified by the spectral density that will be of the form $J(\omega)$,$J_D(\omega)$, $J_1(\omega)$ or $J_2(\omega)$.

To describe the full PSIIRC photocell dynamics under the operation conditions illustrated in Fig. \ref{fig:photocell} we consider two frameworks. \textit{(1) A hybrid framework} that accounts for a non-perturbative approach to the  exciton dynamics using the hierarchical equations of motion\cite{Ishizaki2009,Shi2009,Tanimura2012} in combination with incoherent transfer rates \cite{Kreisbeck2011} to and from all other states. The non-perturbative expansion of the exciton dynamics  is used to account accurately for the effects of $J_D(\omega)$, $J_1(\omega)$ or $J_2(\omega)$. Incoherent rates connecting the exciton states with the rest of states in the photocell are defined using a Lindblad dissipator coupled to each auxiliary density matrix in the expansion \cite{Kreisbeck2011}. Supplementary Note 2 presents further details of the hierarchical expansion of exciton dynamics under this scheme. Converged dynamics are obtained by terminating the hierarchical expansion at level $N=8$ for $J_D(\omega)$ and level $N=5$ for $J_1(\omega)$ and $J_2(\omega)$. Only the $K=0$ Matsubara term was explicitly accounted for, though a Markovian truncation term for Matsubara frequencies was included to capture some finite temperature effects\cite{Shi2009}.

\textit{(2) A Pauli master equation} for electronic state populations is also considered, similarly to the approach followed on Ref. \cite{Novoderezhkin2011}. The Pauli rate equations have the form
%\begin{equation}
$|\dot{P}\rangle\rangle = M |P\rangle\rangle$,
%\end{equation}
where $|P\rangle\rangle$ is a vector of state populations in the basis $\{ |g\rangle,\, |X_1\rangle,\,\cdots |X_6\rangle,\,|I\rangle,\, |\alpha \rangle,\, |\beta \rangle \}$ and $M$ is a stochastic matrix containing the rates for transfer between these electronic states. Modified Redfield theory as presented in Supplementary Note 2 is used to compute population transfer among exciton states\cite{Yang2002,Zhang1998}. In both frameworks (1) and (2) we assume weak and incoherent coupling from excited states to the primary CT state as well as weak coupling between charge transfer states. The transfer from exciton states $|X_n\rangle$ to the intermediate CT $|I\rangle$ are given by Generalised F\"{o}rster theory, and F\"{o}rster-like rates are used to describe transfer between the CT states $|I\rangle$ and $|\alpha \rangle$ \cite{OReilly2014}. Other incoherent rates are described in Supplementary Note 2. Theoretical validity of this framework is discussed in Supplementary Note 3 along with a systematic comparison of the predictions of frameworks (1) and (2).

\subsection*{Theory of full counting statistics}
We envisage our photocell positioned between source and drain leads which supply or remove electrons from the system respectively. The leads are taken as weakly coupled fermionic reservoirs in the limit of infinite bias, such that their influence is described by Lindblad-type dissipators \cite{Harbola2006}, as specified in Supplementary Note  5. With weak coupling to the leads, the theory of full counting statistics\cite{Bagrets2003,Emary2007,Marcos2010} provides a framework to investigate the cumulants of the current passing through the system. This framework is applied to both the hybrid, non-perturbative approach,  and the approximate Pauli model we use to describe the dynamics. For both models a time-local master equation $\dot{\sigma}(t) = \mathcal{M}\sigma(t)$ can be constructed, where the state of the system $\sigma(t)$ is propagated through time by an operator $\mathcal{M}$. This dynamical equation is augmented by a counting field $\chi$, used to single out the incoherent transition across which the electron statistics is counted. For our photocell this is the transition from state $|\alpha\rangle$ to state $|\beta\rangle$ where an electron is transferred to the drain lead. This leads to the time propagator $\mathcal{M}(\chi) = \mathcal{M}_0 + e^{i\chi}\mathcal{M}_J$ where $\mathcal{M}_0$ describes the time evolution of the system between counting events and $\mathcal{M}_J$ is the jump matrix describing hopping events between the system and the drain lead. Further details on the calculation of non-perturbative electron counting statistics using the hierarchical equations of motion \cite{Stones2017} are given in Supplementary Note 5. The zero-frequency cumulants are encoded in the probability distribution of the number of electrons that hop into the drain lead in some long time period\cite{Marcos2010}. A recursive scheme is then followed which generates zero-frequency current cumulants up to any order \cite{Flindt2010} and expresses them in terms of the jump matrix $\mathcal{M}_J$ and the pseudo-inverse $\mathcal{R}$ of the time propagator. The mean $\langle I^1\rangle$ and noise $\langle I^2\rangle$ are given by
\begin{eqnarray}
\langle I^1\rangle &=& \langle\langle \tilde{0}| \mathcal{M}_J |0\rangle\rangle \\
 \langle I^2\rangle &=& \langle\langle \tilde{0}| \mathcal{M}_J - 2\mathcal{M}_J\mathcal{R}\mathcal{M}_J |0\rangle\rangle
\end{eqnarray}
where $\langle\langle \tilde{0}|$ and $|0\rangle\rangle$ are the left and right steady state eigenvectors of the time propagator.

%%%END OF MAIN TEXT%%%

\section*{Acknowledgements}
The authors would like to thank Clive Emary, Elisabet Romero, Vladimir Novoredezkin and Jeroem Elzerman for helpful discussions. Financial support from the Engineering and Physical Sciences Research Council (EPSRC UK) Grant EP/G005222/1 and from the EU FP7 Project PAPETS (GA 323901) is gratefully acknowledged.

\section*{Author contributions statement}
A.O-C designed the research, R.S and H.H-N carried out the calculations. RS, H.H-N, RvG and A.O-C analysed the results and wrote the manuscript.

\section*{Additional information}
The authors declare no competing financial interests.


%\balance


%%%REFERENCES%%%
\providecommand*{\mcitethebibliography}{\thebibliography}
\csname @ifundefined\endcsname{endmcitethebibliography}
{\let\endmcitethebibliography\endthebibliography}{}

\end{document}